\documentclass[sigconf]{acmart}

\usepackage{booktabs}
\usepackage{multirow}
\usepackage{caption}
\usepackage{graphics}
\usepackage{enumitem}
\usepackage{hyperref}
\usepackage{bm}

\setcopyright{none}

\copyrightyear{2022}
\acmYear{2022}
\setcopyright{acmcopyright}\acmConference[ASE '22]{37th IEEE/ACM International Conference on Automated Software Engineering}{October 10--14, 2022}{Rochester, MI, USA}
\acmBooktitle{37th IEEE/ACM International Conference on Automated Software Engineering (ASE '22), October 10--14, 2022, Rochester, MI, USA}
\acmPrice{15.00}
\acmDOI{10.1145/3551349.3561168}
\acmISBN{978-1-4503-9475-8/22/10}

\begin{document}
\newcommand{\STAB}[1]{\begin{tabular}{@{}c@{}}#1\end{tabular}}

\author{Anh T. V. Dau*}
\author{Thang Nguyen-Duc}
\author{Hoang Thanh-Tung}
\email{{anhdtv7, thangnd34, tunght18}@fsoft.com.vn}
\affiliation{%
  \institution{FPT Software AI Center, Viet Nam}
   \country{}
}



\author{Nghi D. Q. Bui*}
\email{dqnbui.2016@smu.edu.sg}
\affiliation{%
  \institution{School of Information Systems \\ Singapore Management University}
   \country{}
  }



\title{Towards Using Data-Influence Methods to Detect Noisy Samples in Source Code Corpora}

\begin{abstract}

Despite the recent trend of developing and applying neural source code models to software engineering tasks, the quality of such models is insufficient for real-world use. This is because there could be noise in the source code corpora used to train such models.
We adapt \textit{data-influence methods} to detect such noises in this paper. Data-influence methods are used in machine learning to evaluate the similarity of a target sample to the correct samples in order to determine whether or not the target sample is noisy. Our evaluation results show that data-influence methods can identify noisy samples from neural code models in classification-based tasks. This approach will contribute to the larger vision of developing better neural source code models from a \textit{data-centric} perspective, which is a key driver for developing useful source code models in practice.

\end{abstract}

\maketitle
	\def\thefootnote{*}\footnotetext{Equal Contributions}\def\thefootnote{\arabic{footnote}}

\section{Introduction}
\label{sec:intro}

Research in the area of \textit{Deep Learning for Code} ~\cite{liu2021deep, li_gated_2015, allamanis2016convolutional, watson2020systematic, DBLP:journals/corr/abs-1711-00740,DBLP:conf/aaai/MouLZWJ16,alon2019code2vec} mostly relies on a large code corpus of code that
allows deep learning methods to reason about source code properties. 
However, the real-world usage of such code models is still limited due to their quality. 
We observe that the source code data used for training code models is collected using a variety of heuristics, such as commit messages, tags provided by code competition websites ~\cite{husain2019codesearchnet, gu2018deep} and people tend to assume that the label of such data is accurate, despite the fact that it may contain a lot of noise ~\cite{husain2019codesearchnet}. 
There are many kinds of noise for classification-based tasks,
but in our work, mislabeled examples are referred to as noisy examples.

We find that the majority of code learning research focuses on improving performance from a \textit{model-centric} standpoint. 
This means that the dataset will remain constant while new models are being proposed to improve performance. 
There are recent efforts to analyze or propose methods to create high-quality datasets for software engineering~\cite{khan2020impact,shome2022data,sun2022importance, zhao2021impact, liu2021opportunities} from the \textit{data-centric} standpoint. 
In the data-centric approach, the model remains fixed while the quality of the datasets used to train such model gets improved. 
\cite{khan2020impact,shome2022data} used simple rules to filter noise.
Rule-based methods do not scale well to large and complex datasets.
\cite{sun2022importance} proposed a learning-based approach for measuring the alignment between code and text for code search data.
In this paper, we approach the problem by adapting data-influence methods \cite{pezeshkpour-etal-2021-empirical}
to detect noises and propose some strategies to enhance the quality of datasets.
By identifying noisy samples, we hope to improve the source code model's performance. 
Results from learning theory also suggest that models trained on clean datasets converge faster and are more robust \cite{ng_2022}.
Data influence methods calculate the influence of training examples on model predictions. 
They track the changes in loss at test data points whenever the training example of interest is used. 
Finally, these methods will provide an influence score. 
This score indicates whether the sample is noisy or not. 

In this work, we concentrate on classification-based tasks, including code classification and defect prediction. For these two tasks, our evaluation shows that the data-influence methods can successfully identify a large number of noisy samples in the training dataset. Furthermore, retraining the models on good training data improves the models' performance and robustness. 

\smallskip

\smallskip
\section{Technical Details}
\label{sec:method}
\label{sec:approaches}
A source code model trained on a training set $\boldsymbol{Z_{train}}$ is denoted as $\boldsymbol{M}(\cdot; \bm \theta)$ . Assuming $\boldsymbol{Z_{train}}$ contains some noise, our goal is to identify the set of noisy samples $\boldsymbol{Z_{noise}} \subset \boldsymbol{Z_{train}}$. 
By removing $\boldsymbol{Z_{noise}}$, we get a new training set $\boldsymbol{Z_{clean}}$. 
Retraining $\boldsymbol{M}$ on $\boldsymbol{Z_{clean}}$ results in a new model $\boldsymbol{M}_{clean}$.
We use the validation set $\boldsymbol{Z_{val}}$ as the anchor to detect noise and select from $\boldsymbol{Z_{val}}$ a set of correctly labeled samples by using $\boldsymbol{M}$. 
A sample is considered correct if the prediction from the model $\boldsymbol{M}$ match its label with high confidence.
We call this set of correctly labeled samples $\boldsymbol{Z_{gold}}$.

Now, we introduce the data-influence methods. We focus on Influence Function (IF) \cite{koh2017understanding} and TracIn \cite{ pruthi2020estimating} as they are currently state-of-the-art techniques. IF \footnote{TracIn \cite{pruthi2020estimating} follows the same principle. Readers are encouraged to read the original paper to check the details of TracIn's formula.} \cite{koh2017understanding} estimates the influence of a sample $\boldsymbol{Z_{train}}^{(i)}$ on the model $\boldsymbol{M}$ by measuring the influence score in accordance with the change of the loss at $\boldsymbol{Z_{gold}}^{(j)}$ when removing a training sample $\boldsymbol{Z_{train}}^{(i)}$ from the training set. 

We then present the pipeline for our evaluation with data-influence methods as the key component to detect noisy samples.
\begin{enumerate}[leftmargin=*]
	\item Initially, we train the model on training set $\boldsymbol{Z_{train}}$, result in model $\boldsymbol{M}$. We use $\boldsymbol{M}$ 
	to randomly select $N$ correctly predicted samples from the validation set $\boldsymbol{Z_{val}}$, resulting in $\boldsymbol{Z_{gold}}$.
	
	\item For each sample in 
	$\boldsymbol{Z_{train}}$, we compute the influence score with all samples in $\boldsymbol{Z_{gold}}$. The score for each training sample is: $S(\boldsymbol{Z_{train}}^{(i)}, \boldsymbol{Z_{gold}} ) = \sum_{j=1}^{N}  S(\boldsymbol{Z_{train}}^{(i)}, \boldsymbol{Z_{gold}}^{(j)})$.
	A negative score shows that $\boldsymbol{Z_{train}}^{(i)}$ has bad influence on the model $\boldsymbol{M}$, which means that $\boldsymbol{Z_{train}}^{(i)}$ is likely to be mislabeled.
	
	\item The top $k\%$ samples with the lowest score, denoted as $\boldsymbol{Z_{noise}}$, are deemed to be noisy. We then remove $\boldsymbol{Z_{noise}}$ from $\boldsymbol{Z_{train}}$ to create the new training set $\boldsymbol{Z_{clean}}$. On $\boldsymbol{Z_{clean}}$, the model is retrained from scratch using the same hyperparameters.
\end{enumerate}

\smallskip
\section{Evaluation}
\label{sec:eval}
We introduce the datasets and source code models used in our experiments.

\textbf{Datasets:} Two types of dataset are involved in our evaluation:
\begin{table}[t]
    \small
	\caption{Results on identifying the mislabeled samples on Synthetic Noisy dataset.}
	\label{tab:acc_tonghop}
	\fontsize{7.0}{9.5}\selectfont 
    \begin{tabular}{@{}l|l|c|c|c@{}}
    \toprule
                              & Method & $k=1$                                 & $k=5$            & $k=10$           \\ \midrule
    \multirow{2}{*}{ASTNN}    & IF     & 94.20 $\pm$ 3.72                      & 84.88 $\pm$ 2.26 & 59.71 $\pm$ 0.32 \\
                              & TracIn & 91.09 $\pm$ 5.06                      & 79.96 $\pm$ 1.29 & 55.97 $\pm$ 1.14 \\ \midrule
    \multirow{2}{*}{CodeBERT} & IF     & \multicolumn{1}{l|}{64.15 $\pm$ 1.07} & 31.50 $\pm$ 1.28 & 14.92 $\pm$ 1.33 \\
                              & TracIn & \multicolumn{1}{l|}{72.36 $\pm$ 2.39} & 49.30 $\pm$ 1.33 & 35.36 $\pm$ 1.15 \\ \bottomrule
    \end{tabular}
\end{table}

\begin{enumerate}[leftmargin=*]
    \item \textbf{Synthetic Noisy Dataset}:
    We inject random noise to a clean dataset. We chose the POJ-104~\cite{mou2016convolutional}, a commonly used dataset for code classification. This dataset contains 52,000 C programs divided into 104 classes of 500 programs each.
    We randomly select 10\% samples in each class and randomly relabel these samples. Now we have a training set $\boldsymbol{Z_{train}}$ containing both clean and noisy data.
    \item \textbf{Real Noisy Dataset}: As shown in ~\cite{khan2020impact}, the dataset used in defect prediction task might contain a significant amount of noise. 
    We chose Devign~\cite{Devign} as the representative after confirming that there are noises in the dataset. Devign dataset includes 21,854 potententially vulnerable C functions collected from open source projects. Each function is manually labeled by software security experts as vulnerable or not.
\end{enumerate}


\textbf{Source code Models: } We take the public code artifacts from ASTNN\footnote{https://github.com/zhangj111/astnn}, CodeBERT\footnote{https://github.com/microsoft/CodeBERT} to reproduce results reported in the original works.

\subsection{Evaluation Results}
\label{sec:real_results}

Firstly, we evaluate our method on the synthetic noisy dataset. Table ~\ref{tab:acc_tonghop} shows the performance of IF and TracIn in identifying the noisy examples on $\boldsymbol{Z_{train}}$. 
In practice, we do not know how many percent of the samples in the dataset are noisy, so we choose different values of $k$. With ASTNN, in top $k=10\%$ samples with the lowest score, both IF and TracIn can detect more than 55\% noisy samples, and when $k=1\%$, more than 91\% samples in this subset are noise. 
Next, we evaluate data-influence methods on real noisy dataset. 
Column \textit{Test ACC} in table ~\ref{tab:real_dataset} shows the performance of models trained on the original dataset. We then calculate the score for all training instances with the same procedure in our evaluation pipeline. The top $1\%$ (best hyperparameter) samples with the lowest score are selected. We also include Random - a baseline randomly selecting a $1\%$ sample set as $\boldsymbol{Z_{noise}}$.
The results in table ~\ref{tab:real_dataset} show that by using data-influence methods, there are improvements in terms of ACC after retraining. 
In general, the results in Table ~\ref{tab:acc_tonghop} and Table ~\ref{tab:real_dataset} support our hypothesis that data-influence methods are effective at detecting noise in code corpus; and removing noises and re-training with cleaner datasets improves the performance of code models 


\begin{table}
	\caption{Results after retraining on the Devign dataset.}
	\label{tab:real_dataset}
	\fontsize{7.0}{11}\selectfont 
    \begin{tabular}{@{}c|c|c|c@{}}
    \toprule
    \multicolumn{1}{l|}{}     & \multicolumn{1}{l|}{Test ACC}     & \multicolumn{1}{l|}{Method} & \multicolumn{1}{l}{Test ACC after removing} \\ \midrule
    \multirow{3}{*}{CodeBERT} & \multirow{3}{*}{62.91 $\pm$ 0.08} & IF                          & \textbf{63.31 $\pm$ 0.10}  \\
                              &                                   & TracIn                      & \textbf{63.40 $\pm$ 0.20}  \\
                              &                                   & Random                      & 61.73 $\pm$ 0.05                            \\ \bottomrule
    \end{tabular}
	\smallskip
\end{table}
\vspace{-7pt}
\smallskip

\section{Discussion \& Conclusion}
\label{sec:diss}
We present a novel data-centric perspective for enhancing the quality of source code models by using data-influence methods. We performed various analyses on several baselines and obtained potentially promising results for improving the quality of source code data. There are numerous aspects that we can investigate in the future. We mostly rely on synthetic noisy datasets to perform the evaluation. Also, we only concentrate on classification-based tasks while we can do the same for many other tasks.For example, when performing a generation-based task like code summarization, the comments and method body are extracted from code snippets collected on Github. However, not all of the developers' comments reflect the functionality of the given code snippet; this can also be interpreted as noise and should be carefully examined too. In the future, we intend to pursue our research in three directions: (1) Identifying more noisy datasets to analyze and providing insights on the noises of such datasets; (2) Improving the methods to detect noisy data; and (3) Applying the methods to a broader range of software engineering tasks, such as code summarization, bug detection, and code translation.


\section{Acknowledgements}
\label{sec:acl}
This work is partly funded by FPT Software AI Center. We also thank the anonymous reviewers for their insightful comments and suggestions.
	

\bibliographystyle{ACM-Reference-Format}
\bibliography{reference}


\begin{thebibliography}{20}


\ifx \showCODEN    \undefined \def \showCODEN     #1{\unskip}     \fi
\ifx \showDOI      \undefined \def \showDOI       #1{#1}\fi
\ifx \showISBNx    \undefined \def \showISBNx     #1{\unskip}     \fi
\ifx \showISBNxiii \undefined \def \showISBNxiii  #1{\unskip}     \fi
\ifx \showISSN     \undefined \def \showISSN      #1{\unskip}     \fi
\ifx \showLCCN     \undefined \def \showLCCN      #1{\unskip}     \fi
\ifx \shownote     \undefined \def \shownote      #1{#1}          \fi
\ifx \showarticletitle \undefined \def \showarticletitle #1{#1}   \fi
\ifx \showURL      \undefined \def \showURL       {\relax}        \fi
\providecommand\bibfield[2]{#2}
\providecommand\bibinfo[2]{#2}
\providecommand\natexlab[1]{#1}
\providecommand\showeprint[2][]{arXiv:#2}

\bibitem[Allamanis et~al\mbox{.}(2018)]%
        {DBLP:journals/corr/abs-1711-00740}
\bibfield{author}{\bibinfo{person}{Miltiadis Allamanis}, \bibinfo{person}{Marc
  Brockschmidt}, {and} \bibinfo{person}{Mahmoud Khademi}.}
  \bibinfo{year}{2018}\natexlab{}.
\newblock \showarticletitle{Learning to Represent Programs with Graphs}, In
  \bibinfo{booktitle}{International Conference on Learning Representations
  (ICLR)}.
\newblock \bibinfo{journal}{\emph{CoRR}}.
\newblock
\urldef\tempurl%
\url{https://doi.org/arXiv:1711.00740}
\showDOI{\tempurl}


\bibitem[Allamanis et~al\mbox{.}(2016)]%
        {allamanis2016convolutional}
\bibfield{author}{\bibinfo{person}{Miltiadis Allamanis}, \bibinfo{person}{Hao
  Peng}, {and} \bibinfo{person}{Charles Sutton}.}
  \bibinfo{year}{2016}\natexlab{}.
\newblock \showarticletitle{A convolutional attention network for extreme
  summarization of source code}. In \bibinfo{booktitle}{\emph{International
  conference on machine learning}}. PMLR, \bibinfo{pages}{2091--2100}.
\newblock


\bibitem[Alon et~al\mbox{.}(2019)]%
        {alon2019code2vec}
\bibfield{author}{\bibinfo{person}{Uri Alon}, \bibinfo{person}{Meital
  Zilberstein}, \bibinfo{person}{Omer Levy}, {and} \bibinfo{person}{Eran
  Yahav}.} \bibinfo{year}{2019}\natexlab{}.
\newblock \showarticletitle{code2vec: Learning distributed representations of
  code}.
\newblock \bibinfo{journal}{\emph{Proceedings of the ACM on Programming
  Languages}} \bibinfo{volume}{3}, \bibinfo{number}{POPL}
  (\bibinfo{year}{2019}), \bibinfo{pages}{1--29}.
\newblock


\bibitem[Gu et~al\mbox{.}(2018)]%
        {gu2018deep}
\bibfield{author}{\bibinfo{person}{Xiaodong Gu}, \bibinfo{person}{Hongyu
  Zhang}, {and} \bibinfo{person}{Sunghun Kim}.}
  \bibinfo{year}{2018}\natexlab{}.
\newblock \showarticletitle{Deep code search}. In
  \bibinfo{booktitle}{\emph{2018 IEEE/ACM 40th International Conference on
  Software Engineering (ICSE)}}. IEEE, \bibinfo{pages}{933--944}.
\newblock


\bibitem[Husain et~al\mbox{.}(2019)]%
        {husain2019codesearchnet}
\bibfield{author}{\bibinfo{person}{Hamel Husain}, \bibinfo{person}{Ho-Hsiang
  Wu}, \bibinfo{person}{Tiferet Gazit}, \bibinfo{person}{Miltiadis Allamanis},
  {and} \bibinfo{person}{Marc Brockschmidt}.} \bibinfo{year}{2019}\natexlab{}.
\newblock \showarticletitle{Codesearchnet challenge: Evaluating the state of
  semantic code search}.
\newblock \bibinfo{journal}{\emph{arXiv preprint arXiv:1909.09436}}
  (\bibinfo{year}{2019}).
\newblock


\bibitem[Khan et~al\mbox{.}(2020)]%
        {khan2020impact}
\bibfield{author}{\bibinfo{person}{Shihab~Shahriar Khan},
  \bibinfo{person}{Nishat~Tasnim Niloy}, \bibinfo{person}{Md~Aquib Azmain},
  {and} \bibinfo{person}{Ahmedul Kabir}.} \bibinfo{year}{2020}\natexlab{}.
\newblock \showarticletitle{Impact of Label Noise and Efficacy of Noise Filters
  in Software Defect Prediction.}. In \bibinfo{booktitle}{\emph{SEKE}}.
  \bibinfo{pages}{347--352}.
\newblock


\bibitem[Koh and Liang(2017)]%
        {koh2017understanding}
\bibfield{author}{\bibinfo{person}{Pang~Wei Koh} {and} \bibinfo{person}{Percy
  Liang}.} \bibinfo{year}{2017}\natexlab{}.
\newblock \showarticletitle{Understanding black-box predictions via influence
  functions}. In \bibinfo{booktitle}{\emph{International conference on machine
  learning}}. PMLR, \bibinfo{pages}{1885--1894}.
\newblock


\bibitem[Li et~al\mbox{.}(2016)]%
        {li_gated_2015}
\bibfield{author}{\bibinfo{person}{Yujia Li}, \bibinfo{person}{Daniel Tarlow},
  \bibinfo{person}{Marc Brockschmidt}, {and} \bibinfo{person}{Richard Zemel}.}
  \bibinfo{year}{2016}\natexlab{}.
\newblock \showarticletitle{Gated Graph Sequence Neural Networks}, In
  \bibinfo{booktitle}{International Conference on Learning Representations
  (ICLR)}.
\newblock \bibinfo{journal}{\emph{arXiv:1511.05493 [cs, stat]}}.
\newblock
\newblock
\shownote{arXiv: 1511.05493}.


\bibitem[Liu et~al\mbox{.}(2021b)]%
        {liu2021opportunities}
\bibfield{author}{\bibinfo{person}{Chao Liu}, \bibinfo{person}{Xin Xia},
  \bibinfo{person}{David Lo}, \bibinfo{person}{Cuiyun Gao},
  \bibinfo{person}{Xiaohu Yang}, {and} \bibinfo{person}{John Grundy}.}
  \bibinfo{year}{2021}\natexlab{b}.
\newblock \showarticletitle{Opportunities and challenges in code search tools}.
\newblock \bibinfo{journal}{\emph{ACM Computing Surveys (CSUR)}}
  \bibinfo{volume}{54}, \bibinfo{number}{9} (\bibinfo{year}{2021}),
  \bibinfo{pages}{1--40}.
\newblock


\bibitem[Liu et~al\mbox{.}(2021a)]%
        {liu2021deep}
\bibfield{author}{\bibinfo{person}{Yue Liu}, \bibinfo{person}{Chakkrit
  Tantithamthavorn}, \bibinfo{person}{Li Li}, {and} \bibinfo{person}{Yepang
  Liu}.} \bibinfo{year}{2021}\natexlab{a}.
\newblock \showarticletitle{Deep Learning for Android Malware Defenses: a
  Systematic Literature Review}.
\newblock \bibinfo{journal}{\emph{arXiv preprint arXiv:2103.05292}}
  (\bibinfo{year}{2021}).
\newblock


\bibitem[Mou et~al\mbox{.}(2016a)]%
        {DBLP:conf/aaai/MouLZWJ16}
\bibfield{author}{\bibinfo{person}{Lili Mou}, \bibinfo{person}{Ge Li},
  \bibinfo{person}{Lu Zhang}, \bibinfo{person}{Tao Wang}, {and}
  \bibinfo{person}{Zhi Jin}.} \bibinfo{year}{2016}\natexlab{a}.
\newblock \showarticletitle{Convolutional Neural Networks over Tree Structures
  for Programming Language Processing}. In
  \bibinfo{booktitle}{\emph{Proceedings of the Thirtieth {AAAI} Conference on
  Artificial Intelligence}}. \bibinfo{pages}{1287--1293}.
\newblock


\bibitem[Mou et~al\mbox{.}(2016b)]%
        {mou2016convolutional}
\bibfield{author}{\bibinfo{person}{Lili Mou}, \bibinfo{person}{Ge Li},
  \bibinfo{person}{Lu Zhang}, \bibinfo{person}{Tao Wang}, {and}
  \bibinfo{person}{Zhi Jin}.} \bibinfo{year}{2016}\natexlab{b}.
\newblock \showarticletitle{Convolutional neural networks over tree structures
  for programming language processing}. In \bibinfo{booktitle}{\emph{Thirtieth
  AAAI conference on artificial intelligence}}.
\newblock


\bibitem[Ng(2022)]%
        {ng_2022}
\bibfield{author}{\bibinfo{person}{Andrew Ng}.}
  \bibinfo{year}{2022}\natexlab{}.
\newblock \bibinfo{title}{Andrew Ng "the data-centric AI approach"}.
\newblock
\newblock
\urldef\tempurl%
\url{https://www.youtube.com/watch?v=TU6u_T-s68Y}
\showURL{%
\tempurl}


\bibitem[Pezeshkpour et~al\mbox{.}(2021)]%
        {pezeshkpour-etal-2021-empirical}
\bibfield{author}{\bibinfo{person}{Pouya Pezeshkpour}, \bibinfo{person}{Sarthak
  Jain}, \bibinfo{person}{Byron Wallace}, {and} \bibinfo{person}{Sameer
  Singh}.} \bibinfo{year}{2021}\natexlab{}.
\newblock \showarticletitle{An Empirical Comparison of Instance Attribution
  Methods for {NLP}}. In \bibinfo{booktitle}{\emph{Proceedings of the 2021
  Conference of the North American Chapter of the Association for Computational
  Linguistics: Human Language Technologies}}. \bibinfo{publisher}{Association
  for Computational Linguistics}, \bibinfo{address}{Online},
  \bibinfo{pages}{967--975}.
\newblock
\urldef\tempurl%
\url{https://doi.org/10.18653/v1/2021.naacl-main.75}
\showDOI{\tempurl}


\bibitem[Pruthi et~al\mbox{.}(2020)]%
        {pruthi2020estimating}
\bibfield{author}{\bibinfo{person}{Garima Pruthi}, \bibinfo{person}{Frederick
  Liu}, \bibinfo{person}{Satyen Kale}, {and} \bibinfo{person}{Mukund
  Sundararajan}.} \bibinfo{year}{2020}\natexlab{}.
\newblock \showarticletitle{Estimating training data influence by tracing
  gradient descent}.
\newblock \bibinfo{journal}{\emph{Advances in Neural Information Processing
  Systems}}  \bibinfo{volume}{33} (\bibinfo{year}{2020}),
  \bibinfo{pages}{19920--19930}.
\newblock


\bibitem[Shome et~al\mbox{.}(2022)]%
        {shome2022data}
\bibfield{author}{\bibinfo{person}{Arumoy Shome}, \bibinfo{person}{Luis Cruz},
  {and} \bibinfo{person}{Arie van Deursen}.} \bibinfo{year}{2022}\natexlab{}.
\newblock \showarticletitle{Data Smells in Public Datasets}.
\newblock \bibinfo{journal}{\emph{arXiv preprint arXiv:2203.08007}}
  (\bibinfo{year}{2022}).
\newblock


\bibitem[Sun et~al\mbox{.}(2022)]%
        {sun2022importance}
\bibfield{author}{\bibinfo{person}{Zhensu Sun}, \bibinfo{person}{Li Li},
  \bibinfo{person}{Yan Liu}, \bibinfo{person}{Xiaoning Du}, {and}
  \bibinfo{person}{Li Li}.} \bibinfo{year}{2022}\natexlab{}.
\newblock \showarticletitle{On the importance of building high-quality training
  datasets for neural code search}. In \bibinfo{booktitle}{\emph{Proceedings of
  the 44th International Conference on Software Engineering}}.
  \bibinfo{pages}{1609--1620}.
\newblock


\bibitem[Watson et~al\mbox{.}(2020)]%
        {watson2020systematic}
\bibfield{author}{\bibinfo{person}{Cody Watson}, \bibinfo{person}{Nathan
  Cooper}, \bibinfo{person}{David~Nader Palacio}, \bibinfo{person}{Kevin
  Moran}, {and} \bibinfo{person}{Denys Poshyvanyk}.}
  \bibinfo{year}{2020}\natexlab{}.
\newblock \showarticletitle{A Systematic Literature Review on the Use of Deep
  Learning in Software Engineering Research}.
\newblock \bibinfo{journal}{\emph{arXiv preprint arXiv:2009.06520}}
  (\bibinfo{year}{2020}).
\newblock


\bibitem[Zhao et~al\mbox{.}(2021)]%
        {zhao2021impact}
\bibfield{author}{\bibinfo{person}{Yanjie Zhao}, \bibinfo{person}{Li Li},
  \bibinfo{person}{Haoyu Wang}, \bibinfo{person}{Haipeng Cai},
  \bibinfo{person}{Tegawend{\'e}~F Bissyand{\'e}}, \bibinfo{person}{Jacques
  Klein}, {and} \bibinfo{person}{John Grundy}.}
  \bibinfo{year}{2021}\natexlab{}.
\newblock \showarticletitle{On the impact of sample duplication in
  machine-learning-based android malware detection}.
\newblock \bibinfo{journal}{\emph{ACM Transactions on Software Engineering and
  Methodology (TOSEM)}} \bibinfo{volume}{30}, \bibinfo{number}{3}
  (\bibinfo{year}{2021}), \bibinfo{pages}{1--38}.
\newblock


\bibitem[Zhou et~al\mbox{.}(2019)]%
        {Devign}
\bibfield{author}{\bibinfo{person}{Yaqin Zhou}, \bibinfo{person}{Shangqing
  Liu}, \bibinfo{person}{Jing~Kai Siow}, \bibinfo{person}{Xiaoning Du}, {and}
  \bibinfo{person}{Yang Liu}.} \bibinfo{year}{2019}\natexlab{}.
\newblock \showarticletitle{Devign: Effective Vulnerability Identification by
  Learning Comprehensive Program Semantics via Graph Neural Networks}. In
  \bibinfo{booktitle}{\emph{Advances in Neural Information Processing Systems
  32: Annual Conference on Neural Information Processing Systems 2019, NeurIPS
  2019, December 8-14, 2019, Vancouver, BC, Canada}},
  \bibfield{editor}{\bibinfo{person}{Hanna~M. Wallach}, \bibinfo{person}{Hugo
  Larochelle}, \bibinfo{person}{Alina Beygelzimer}, \bibinfo{person}{Florence
  d'Alch{\'{e}}{-}Buc}, \bibinfo{person}{Emily~B. Fox}, {and}
  \bibinfo{person}{Roman Garnett}} (Eds.). \bibinfo{pages}{10197--10207}.
\newblock
\urldef\tempurl%
\url{https://proceedings.neurips.cc/paper/2019/hash/49265d2447bc3bbfe9e76306ce40a31f-Abstract.html}
\showURL{%
\tempurl}


\end{thebibliography}

\end{document}